\documentclass[11pt]{article} 
 
\usepackage{amssymb,cite} 
\usepackage{graphicx} 
\usepackage{latexsym} 

\usepackage{epsfig,amsfonts} 

\usepackage{bm}
\usepackage[ansinew]{inputenc}
\usepackage{rotating}
\usepackage{euscript}
\newcommand{\bra}{\left \langle}
\newcommand{\ket}{\right \rangle}
\newcommand{\eq}{\begin{equation}} 
\newcommand{\en}{\end{equation}} 

\def\one{{\rm 1\kern -.9mm l}}                             %

\newcommand{\eqa}{\begin{eqnarray}}
\newcommand{\ena}{\end{eqnarray}}
\def\beq{\begin{equation}}
\def\eeq{\end{equation}}
\def\beq{\begin{equation}}
\def\eeq{\end{equation}}
\def\beqa{\begin{eqnarray}}
\def\eeqa{\end{eqnarray}}
\def\de{\partial}

\begin{document}
\begin{titlepage}
\vskip0.5cm
\begin{flushright}
\end{flushright}
\vskip0.5cm
\begin{center}
{\Large\bf A new approach to the study of effective string corrections in LGTs
 } 
\end{center}
\vskip1.3cm
\centerline{
M.~Caselle and  M. Zago}
 \vskip1.0cm
 \centerline{\sl  Dipartimento di Fisica
 Teorica dell'Universit\`a di Torino and I.N.F.N.,}
 \centerline{\sl Via Pietro~Giuria 1, I-10125 Torino, Italy}
 \centerline{{\sl
e--mail:} \hskip 6mm
 \texttt{(caselle)(zago)@to.infn.it}}
 \vskip2.0cm
\begin{abstract}
We propose a new approach to the study of the interquark potential in Lattice Gauge Theories. Instead of looking at the expectation value of 
Polyakov loop correlators we study the modifications induced in the chromoelectric flux by the presence of the Polyakov loops. In abelian LGTs, thanks to
duality, this study can be performed in a very efficient way, allowing to reach high precision at a reasonable
CPU cost. The major advantage of this strategy is that it allows to eliminate the dominant effective string correction to the
interquark potential (the L\"uscher term) thus giving an unique opportunity to test higher order corrections. Performing a set of
simulations in the 3d gauge Ising model we were thus able to precisely identify and measure both the quartic and the sextic effective
string corrections to the interquark potential. While the quartic term perfectly agrees with the Nambu-Goto one 
the sextic term is definitely different. Our result seems to disagree with the recent proof of the universality of the sextic correction.
We discuss a few possible explanations of this disagreement.
\end{abstract}
\end{titlepage}

\setcounter{footnote}{0}
\def\thefootnote{\arabic{footnote}}

\section{Introduction}
\label{introsect}
An increasing amount of numerical results supports the idea that the flux tube joining the quark-antiquark pair in 
the confining regime of a gauge theory should be described by an effective string theory.
Starting from the seminal paper by 
L\"uscher, Symanzik and Weisz \cite{Luscher:1980fr} many theoretical efforts have been devoted in these thirty years to 
understand the nature of such string, constrain its action and predict 
its effects on the interquark potential. At the same time, the impressive development, both from the theoretical and from the computational point of view 
of Lattice Gauge Theories (LGTs) provided a better and better numerical 
laboratory to test these predictions (for un updated review of lattice results see: \cite{Teper:2009uf}).
One of the most interesting recent results in this context  is that the first few terms in the perturbative expansion of the string contribution to the interquark potential 
should be universal and thus can be evaluated assuming the simplest possible string action, i.e. the Nambu-Goto string. Universality up to the quartic
order was proved a few years ago by Lu\"scher and Weisz~\cite{lw04} and recently extended to the sixth 
order by Aharony and collaborators \cite{Aharony:2009gg,ak,af} (see also \cite{Dass}). These
results are based on a set of assumptions (which we shall discuss in more detail below) and it would 
be of great importance to test them with a numerical simulation. This is a very difficult task for at least two reasons: 
\begin{itemize} 
\item
 In the standard "zero temperature" interquark potential, higher order corrections are proportional to higher powers in $1/R$ 
 (where $R$ is the interquark distance) and are thus visible only at very short distance where the effective string picture breaks down and spurious effects (due for
 instance to boundary terms) and perturbative contributions become important.
\item
 The dominant string correction (the so called L\"uscher term)  may shadow the subleading terms
 \end{itemize}

In order to overcome these two problems we propose the following strategy. First, instead of working at zero temperature we shall study the 
interquark potential at finite temperature (just below the deconfinement transition). It is well known that in this regime the string corrections  are proportional to $R$
and act as a temperature dependent renormalization of the string tension. In this regime higher order corrections correspond to higher powers of $T$ and can be observed much
better than in the zero temperature limit.
Second, in order to eliminate the dominant L\"uscher term we shall not measure directly
 the interquark potential, but shall instead study the changes induced in 
the flux configuration by the presence of the Polyakov loops. 
We shall show below that 
as a consequence of this choice the L\"uscher term vanishes. This makes this observable an
unique tool to explore higher order corrections. 
Another reason of interest of 
this approach is that it is deeply related to another important issue of the effective string description of LGTs, i.e.
the study of the flux tube thickness. This observable has
recently attracted much interest~\cite{Gliozzi:2010zv,Allais:2008bk}. 
In particular it has been shown that in the high temperature regime in which we are
presently interested the width of the flux tube increases linearly with the interquark distance and not logarithmicaly as one would
naively expect~\cite{Gliozzi:2010zv,Allais:2008bk}. This linear increase is  related to the linear increase in the flux energy that we observe 
here. In both cases the slope is temperature dependent and contains information on the higher order effective string
corrections in which we are intersted.

This paper is organized as follows. We shall introduce our approach and define the flux observables in 
which we are interested in sect.2 . Sect.3 will be devoted to a detailed discussion of the
effective string predictions for these observables. In sect.4 we shall compare these predictions with a set of high precision simulations
performed in the 3d gauge Ising model while the last section will be devoted to some concluding remarks and a discussion of the
interplay of our results with the universality of the sixth order correction recently discussed by  
Aharony and collaborators in \cite{Aharony:2009gg,ak,af}.

\section{Flux density in presence of the Polyakov loops.}

The lattice operator which measures the flux through a plaquette $p$ in presence of two Polyakov loops $P$, $P'$ is:
\eq 
\bra\phi(p;P,P')\ket=\frac{\bra P P'^\dagger~U_p\ket}{\bra PP'^\dagger \ket}-\bra U_p\ket 
\label{flux2} 
\en 
where $U_p$ is, as usual, the trace of the orderd product of link variables along the plaquette $p$.
Different possible orientations of the plaquette measure different components of the flux.
We shall be interested in the following in the mean flux density, i.e. the sum of $\phi(p;P,P')$ over all the positions and orientations of the plaquettes, normalized to
the number of plaquettes of the lattice. 
Due to translational invariance this quantity will depend only on the distance $R$ between the two loops and on the inverse temperature $L$ (i.e. the lattice size in the
compactified direction). 
Let us define
\eq
\bra\Phi(R,L)\ket=\frac{1}{N_p}\sum_p \frac{\bra P P'^\dagger~U_p\ket}{\bra PP'^\dagger \ket} - \bra U_p\ket
\en
where $N_p$ denotes the number of plaquettes of the lattice. 
We shall denote in the following with $N_s$ and $L$ the number of lattice spacings in the spacelike and timelike
directions respectively so that in $d=2+1$ we have $N_p=3N_s^2L$

It is easy to see that if we define:
\eq
Z(R,L)=\langle P^\dagger (R) P(0) \rangle = \frac{1}{Z}\sum_c P^\dagger (R) P(0) ~ e^{\beta \sum_p U_p}  
\label{neq1}
\en
where the partition function in eq.(\ref{neq1}) is given as usual by
\eq
Z=\sum_c ~ e^{\beta \sum_p U_p}  
\en
and $\sum_c$ denotes the sum over all the configurations,
then
$\bra\Phi(R,L)\ket$ can be written as:
\eq
\bra\Phi(R,L)\ket=\frac{1}{N_p}\frac{d}{d \beta} log Z(R,L)~~~.
\label{defPhi}
\en
If we neglect for the moment effective string corrections and keep only the area term in $Z$, i.e. $Z(L,R)\sim e^{-\sigma RL}$
we find a linearly rising behaviour for $\bra\Phi(R,L)\ket$:
\eq
\bra\Phi(R,L)\ket=\alpha R
\label{Phi2}
\en
with an angular coefficient 
\eq
\alpha=-\frac{L}{N_p}\frac{d \sigma}{d \beta} 
\label{defalpha}
\en
which does not depend on the finite temperature $1/L$ (which cancels out in the ratio $L/N_p$)  

If we now look at the effective string corrections to this result we immediately observe a very interesting feature of $\bra\Phi(R,L)\ket$ i.e. the fact that 
the dominant term (the so called L\"uscher term) vanishes.

In fact one of the main assumptions
of the effective string approach to LGTs is that the only dimensional parameter which appears in the action\footnote{In principle one could also add to the action a
"boundary term" proportional to a dimensional parameter $b$. This term vanishes at the first order in (2+1) dimensions~\cite{lw04} but it may
appear at higher orders. We shall discuss below this issue.} 
is the string tension $\sigma$. Therefore 
observables like $Z(R,L)$ may depend on $\beta$ 
only through the string tension $\sigma$. This means that 
in the effective string framework we can convert the $\beta$ derivative into a derivative with respect of $\sigma$.

However it is well known~\cite{Luscher:1980fr} that the dominant correction  must be scale invariant and thus $\sigma$ independent. This means that in eq.(\ref{defPhi})
this term disappears and the effective string correction to $\bra\Phi(R,L)\ket$ starts from the first subleading term: the "quartic" 
correction that we shall discuss below.

\section{Effective string corrections}
\label{sect1}

In this section we shall first briefly recall a few general results on the Nambu-Goto string and on its perturbative expansion (a detailed derivation can be found in
\cite{Aharony:2009gg} and\cite{chp03}) and then use these results to 
discuss, in the framework of the Nambu-Goto effective string, how the flux distribution is modified by the presence of the Polyakov loops.

\subsection{The Nambu-Goto string} 

The   action of the Nambu-Goto model is proportional to the area of the string 
world-sheet:
\eq
S=\sigma\int_0^{L}d\tau\int_0^{R} d\varsigma\sqrt{g}\ \ ,\label{action}
\en
where $g$ is the determinant of the two--dimensional metric induced on
the world--sheet by the embedding in $R^d$  
and $\sigma$ is the string tension,
which appears as the only parameter of the effective model.
\par
Eq.~(\ref{action}) is invariant with respect to reparametrization and Weyl
transformations, and a possible choice for quantization of the effective model
is the ``physical gauge''  in which $g$ becomes a function of the transverse displacements of the string world-sheet only. The 
latter (which can be denoted as $X^i(\tau,\varsigma)$) are required to satisfy the boundary conditions relevant to the problem. In the present case, the effective 
string world-sheet associated with a two-point Polyakov loop correlation function obeys periodic b.c. in the compactified direction and Dirichlet b.c. along the 
interquark axis direction:
\eq
X^i(\tau+L,\varsigma)=X^i(\tau,\varsigma); \hskip 1cm X^i(\tau,0)=X^i(\tau,R)=0\ \ .
\en
This gauge fixing implicitly assumes that the world-sheet surface is a single-valued function of $(\tau,\varsigma)$, i.e. overhangs, cuts, or disconnected 
parts are excluded. 
It is well known that rotational symmetry of this model is broken at the quantum level because of the Weyl anomaly, unless $d=26$. 
However this anomaly is known to vanish at large distances \cite{olesen}, and this suggests to use the ``physical gauge'' for
an IR, effective string description also for $d \neq 26$.

Here and in the following, we restrict our attention to the $d=2+1$ case, which is particularly simple, as there is only one transverse degree of freedom ($X$).
In the physical gauge, eq. (\ref{action}) takes the form: 
\eq
S[X]=\sigma\int_0^L d \tau \int_0^R d \varsigma
\sqrt{1+(\de_\tau X)^2+(\de_\varsigma X)^2} \;\;.
\label{squareroot}
\en

This action describes a non-renormalizable, interacting 
QFT in two dimensions; the associated partition function is expected to encode an effective description for this sector of the gauge theory, 
providing a prediction for the VEV of the two-point Polyakov loop correlation function: 
\eq
\label{partitionfunction}
\langle P^\dagger (R) P(0) \rangle = Z(L,R) = \int \mathcal{D} X e^{-S[X]} \;\;.
\en

\subsection{Perturbative expansion of the partition function}

The effective model can be studied perturbatively, 
expanding the square root appearing in eq.~(\ref{squareroot}) in powers of the dimensionless quantity $(\sigma RL)^{-1}$: this approach is expected 
to be consistent with the fact that the effective theory holds in the IR regime of the confined phase. We find:

\eq
Z(L,R)=e^{-\sigma RL} \cdot Z_{1} \cdot \left( 1+ \frac{F_4}{\sigma R^2} + \frac{F_6}{(\sigma R^2)^2} +\cdots \right)
\label{zexp}
\en
where the indices in $F_4$ and $F_6$ recall the fact that they are obtained from the quartic and sextic terms in the expansion of the Nambu-Goto action respectively.

The leading order of this expansion: $Z_1$  corresponds to the partition function of a free boson in two dimensions:
\eq
Z_{1}=\frac{1}{\eta(iu)} 
\label{z1}
\en
where $u=\frac{L}{2R}$ and $\eta$ denotes the  Dedekind  function:
\eq
\eta(\tau)=q^{1\over24}\prod_{n=1}^\infty(1-q^n)\hskip0.5cm
;\hskip0.5cmq=e^{2\pi i\tau}~~~,
\label{eta}
\en
This correction is universal and does not depend on the string tension $\sigma$. Thus, as anticipated in the introduction, it will not contribute 
to $\bra\Phi(R,L)\ket$.

The next to leading corrections was evaluated in~\cite{df83} and can be written as a combination of Eisenstein functions.
\eq
F_4=\frac{ \pi^2 L}{1152 R} \left[ 2E_4 \left( i u \right) - E_2^2 \left( i u \right) \right]  ~~~, 
\label{znlo}
\en
where $E_2$ and $E_4$ are defined as:
\eqa
E_2(\tau)&=&1-24\sum_{n=1}^\infty \sigma_1(n) q^n \label{eisenstein2}\\
E_4(\tau)&=&1+240\sum_{n=1}^\infty \sigma_3(n) q^n \label{eisenstein4}\\
q&=& e^{2\pi i\tau} \;\;, 
\ena
where $\sigma_1(n)$ and $\sigma_3(n)$ are, respectively, the sum of all
divisors of $n$ (including 1 and $n$), and the sum of their cubes.

This expression looks complicated, but it simplifies in the large $R/L$ limit (i.e. $q \to 1$) in which we are interested.
By performing a modular transformation we can rewrite the above expression in terms of $\tilde q \equiv  e^{-4 \pi \frac{R}{L} }$.
In the large $R/L$ limit we may neglect all higher order terms in the $\tilde q$ expansion\footnote{We shall make this statement more rigorous in the section devoted to the
numerical simulations.}
and after some algebra we end up with the following contribution
to $\bra\Phi(R,L)\ket$:
\eq
\bra\Phi(R,L)\ket~=~
\alpha 
\left(\frac{\pi^2}{72\sigma^2 L^4} R  + \frac{\pi}{12\sigma^2 L^3} + \frac{1}{8\sigma^2 L^2}\frac{1}{R}\right) 
\en
We see that as a consequence of the flux tube fluctuations a $1/R$ term appear in the mean flux density and that the linerly rising term of eq.(\ref{Phi2}) gets a
correction proportional to $1/L^4$ and thus becomes temperature dependent.

Higher order terms, like $F_6$, could be derived, in principle, along the same lines  
but, if one is interested in the
Nambu-Goto action only, they can be more simply obtained by expanding in power of $1/(\sigma L R)$ 
the exact solution obtained a few years ago by L\"uscher and Weisz~\cite{lw04} 
and then derived in the covariant formalism in~\cite{Billo:2005iv}. 

In $d=2+1$ dimensions one finds that $Z(R,L)$ is given by a tower of $K_0$ Bessel functions\cite{lw04,Billo:2005iv}: 
\eq 
Z(R,L)= \left\langle P(0,0) P(0,R)\right\rangle 
  =\sum_{n=0}^{\infty}c_n 
  K_{0}({E}_nR). 
\label{zeq1}
\en 
 
where $E_n$ are the closed string energy levels: 
  
 \eq 
  {E}_n=\sigma L
  \left\{1+{8\pi\over\sigma L^2}\left[-\frac{1}{24}\left(d-2\right)+n\right] 
  \right\}^{1/2}. \label{zeq2}
  \en 
 and $\sigma$ denotes the zero temperature string tension.

It is easy to see that in the large $R$ limit only the lowest state $(n=0)$ survives\footnote{This statement is equivalent to the choice of neglecting terms proportional
to  $\tilde q$ that we made in the previous section.} and
 we end up with a single 
$K_0$ function: 
\eq 
\lim_{R\to\infty}  \left\langle P(0,0) P(0,R)\right\rangle   
= c_0  K_{0}({E}_0R). 
\label{rtoinfty} 
\en 

with 
\eq 
  E_0=\sigma L 
  \left(1-{\frac{\pi}{3\sigma L^2}} 
  \right)^{1/2}. 
\label{sigmal}
  \en 

and

\eq
c_0=\frac{L}{2} \sqrt{\frac{\sigma}{\pi}}
\label{zeq3}
\en

From eqs.(\ref{rtoinfty},\ref{sigmal} and  \ref{zeq3})
we find

\eq
\frac{d \log Z(R,L)}{d \beta}=\left(\frac{1}{2\sigma} +
R \frac{K'_0}{K_0} \frac{d E_0}{d\sigma}\right)\frac{d \sigma}{d\beta}
\en

where we used the fact that $Z(R,L)$ is a function of  $\beta$ only through the string tension $\sigma$ 

Using eq.(\ref{sigmal}) we find

\eq
\frac{d E_0}{d\sigma}=\frac{L(1-x/2)}{\sqrt{1-x}}
\en
where we shall denote from now on 
\eq
x\equiv\frac{\pi}{3\sigma L^2}
\label{defx}
\en

Recalling that 
$$K_0'=-K_1$$
and using the asymptotic expansion of the modified Bessel functions
we obtain
\eq
\frac{K_1(E_0R)}{K_0(E_0R)}=1+\frac{1}{2E_0R}-\frac{1}{8(E_0R)^2}+ ...
\en

Collecting all the terms together
we find

\eq
\bra\Phi(R,L)\ket~=~
\alpha 
\left(R A(x) + B(x) + \frac{C(x)}{R} + ...\right) 
\en
where we  defined:
\eq
A(x)=\frac{(1-x/2)}{\sqrt{1-x}}
\label{ng1}
\en
\eq
B(x)=\frac{1}{4\sigma L}\frac{x}{(1-x)}
\en
\eq
C(x)=\frac{1-x/2}{8(L\sigma)^2(1-x)^{3/2}}
\en
and $\alpha$ was defined in eq.(\ref{defalpha}) which we report here for the particular case of a (2+1) dimensional lattice for which
$N_p=3N_s^2 L$:
\eq
\alpha= -\frac{1}{3N_s^2}\frac{d \sigma}{d\beta} ~~.
\label{defalpha3d}
\en

While $\alpha$ depends on the details of the gauge theory  and in particular on the $\beta$ dependence of the 
string tension $\sigma$, the functions $A(x),B(x)$ and $C(x)$ ... encode the information on the effective string model. The particular form listed above for these functions
corresponds to the Nambu-Goto model, however according to~\cite{lw04,Aharony:2009gg} the first two orders in the perturbative expansion in power of $1/\sigma$ (i.e., with
our conventions, the
terms proportional to $1/\sigma^2$ and $1/\sigma^3$)
powers of $x$) should be universal. In order to identify these terms, let us expand in powers of $x$ these functions.

Expanding in $x$ we find
\eq
 A(x)= \left(1+\frac{x^2}{8}+\frac{x^3}{8}+...\right)
\label{def1}
\en
\eq
 B(x)=\frac{1}{\sigma L}\left(\frac{x}{4}+\frac{x^2}{4}+...\right)
\label{def2}
\en
\eq
 C(x)=\frac{1}{8(\sigma L)^2}\left(1+ x+ ...\right)
\label{def3}
\en

which at the first order coincide as expected  with the  corrections obtained with
the previous approach. In addition we find the next to leading corrections of
order $1/\sigma^3$.

\subsection{Boundary corrections}
The analysis of \cite{lw04} excluded the presence of a boundary 
term at the first order (see eq. (2.4) of \cite{lw04}) 
 but could not exclude the presence of boundary terms at higher orders. These higher order boundary terms were recently studied in great detail in~\cite{af}.
 Looking at their result it is easy to see that they cannot contribute to the term proportional to $R$ in the expectation value of $\bra\Phi(R,L)\ket$, 
 but they may affect the following terms (constant and proprotional to $1/R$). For this reason in the following 
 we shall concentrate only on the linear correction.

\section{Numerical simulations}
\label{sect3}

\subsection{Details on the computational setting}
We compared our predictions with a set of high precision simulations in the 3d gauge Ising model.
We performed our simulations using the same methods discussed in \cite{Allais:2008bk,cgmv95}. 
We used duality to map the Polyakov loops correlator
into the partition function of a 3d Ising spin model in which we changed the sign of the coupling 
of all the links dual to the surface bordered by the two Polyakov loops.
We then estimated $\bra\Phi(R,L)\ket$ by simply evaluating the mean energy in presence of these frustrated links. We chose periodic boundary conditions in the original 
gauge Ising model. These b.c. are mapped by duality into a mixture of periodic and antiperiodic b.c. in the dual spin model. However we always chose $N_s$ large enough
to eliminated any contribution from the antiperiodic sectors (which are depressed by terms proportional to $e^{-\sigma N_s L}$ or $e^{-\sigma N_s^2}$). 
It is clear that duality plays a crucial role in this derivation and for this reason we consider the approach discussed in this paper
 as particularly suited for abelian gauge theories, even if in
principle, given enough computational power, it could be extended also to non-abelian models. 

Since, as discussed above, we shall be interested in the following only in the
term proportional to $R$ in $\bra\Phi(R,L)\ket$, (which is the only one which is not affected by boundary corrections) we shall neglect in the following the 
disconnected component $\bra U_p\ket$  of $\bra\Phi(R,L)\ket$.

\subsection{Evaluation of $\alpha$ for the  3d gauge Ising model}

\subsubsection{Naive analysis}
Assuming the scaling behaviour 
\eq
\sigma(\beta)=\sigma_c(\beta_c-\beta)^{2\nu}
\en
we find
\eq
-\frac{1}{\sigma}\frac{d \sigma}{d\beta}=\frac{2\nu}{\beta_c-\beta}
\en
which would lead to a very simple ed elegant form for $\alpha$
\eq
\alpha\equiv \frac{2\nu \sigma}{3N_s^2(\beta_c-\beta)}
\label{naive}
\en

However this is not enough for our purposes.
In order to estimate higher order effective string corrections  we need to evaluate 
the flux density with an uncertainty smaller than 1\% and thus 
it is mandatory to have a rigorous control of all the possible
sources of systematic error up to this threshold. We have in particular four potential sources 
of systematic errors:  the determination of the critical point $\beta_c$ and of the critical index $\nu$, 
the determination of the zero temperature string tension $\sigma$ and 
the presence of scaling corrections which could affect the derivative $\frac{d \sigma}{d\beta}$. Let us addres these point in more detail

\subsubsection{Scaling corrections and critical parameters.}
The most accurate result for the  critical 
temperature of the 3d spin Ising model is $\beta_{c,s}=0.22165455(5)$~\cite{DB03} from which we can immediately obtain $\beta_c$ using the duality relations.
The most accurate values given in the literature for the critical index $\nu$ are $\nu=0.63012(16)$~\cite{phi4pisa} 
from the analysis of high temperature series expansions and  $\nu=0.63002(10)$~\cite{Martin2010}  from a recent finite size scaling 
analysis of Monte Carlo data. We shall adopt this last value in the following analysis. For both $\beta_{\mbox{\tiny{c}}}$ and $\nu$ the available precision is
much smaller than our statistical errors thus we may neglect this possible source of systematic error.
While there are no precise enough results on the scaling behaviour of the string tension we can extract the information that we need from a very precise study
appeared a few years ago~\cite{Caselle:2007yc} on the scaling behaviour of the dual observable: the interface
tension of the 3d Ising spin model. Let us define as $\beta_s$ the dual of the gauge coupling $\beta$ and accordingly as $t=\beta_s-\beta_{c,s}$ 
the reduced temperature of the 3d spin model.

In the neighbourhood of the transition, the interface tension behaves as:
\begin{equation}
\label{powerlaw}
\sigma(\beta) = \sigma_c t^{2\nu} \times (1 + a t^{\theta} + b t  
+ c t^{2 \theta} + d t^{\theta'} + ...) \;\;,
\end{equation}
The most accurate result  for the subleading exponent is $\omega=0.832(6)$~\cite{Martin2010}, 
from which $\theta=\nu \omega=0.5241(33)$. For $\theta'$ we may use the result $\theta'=1.05(7)$~\cite{NeRi84}. 

Since $2 \theta \approx \theta' \approx 1$, we may assume a simplified ansatz for the scaling behaviour:
\begin{equation}
\label{powerfit}
\sigma(\beta) = \sigma_c t^{\mu} \times (1 + a t^{\theta} + b t) \;.
\end{equation}

Using a set of extensive simulations of the interface tension in~\cite{Caselle:2007yc} the authors were able to extract  rather precise
estimates for the free  parameters of the above ansatz:
 $\sigma_c=10.083(8)$, $a=-0.479(26)$ and $b=-2.12(19)$

We are interested in the the value of:
\eq
-\frac{1}{\sigma}\frac{d \sigma}{d\beta}
\en

Let us rewrite it as:

\eq
-\frac{1}{\sigma}\frac{d \sigma}{d\beta}=-\frac{1}{\sigma}\frac{d \beta_s}{d\beta}\frac{d \sigma}{d\beta_s}=
-\frac{1}{\sigma}\frac{d \beta_s}{d\beta}\frac{d \sigma}{d t}
\en
using 
\eq
-\frac{d \beta_s}{d\beta}=\frac{1}{\sinh{2\beta}}
\en

we find, 
\eq
-\frac{1}{\sigma}\frac{d \sigma}{d\beta}=\frac{1}{\sinh{2\beta}}\left(\frac{2\nu}{t}\left[1+\frac{a\theta}{2\nu}t^\theta+\frac{(b-a^2)\theta}{2\nu}t\right] \right)
\en

where we made the approximation $2\theta\sim 1$ which allows us to combine the correction proportional to $a^2$ with the one proportional to $b$. 

In the range of values of $\beta$ that we studied this correction turns out to be larger than our statistical errors and thus must be taken into account

Using the above scaling function we may also obtain very precise estimates for the string tension for
the values of $\beta$ that we studied 

\subsection{Results and comparison with the effective string predictions}

We perfomed simulations for three values of $\beta$: $\beta=0.75180,~0.756427,~0.758266$
which were chosen because for these values the deconfinement temperature is known with very high precision and
coincides with $1/T_c=L_c=8,12,16$ respectively~\cite{Caselle:1995wn}. In particular, we studied several values of $L$ in the 
$\beta=0.75180$ case and a single value of $L$ for $\beta=0.756427,~0.758266$ to test the scaling behaviour of the corrections.
For each value of $L$ we studied a wide set of values of the interquark distance in the range $\frac{1}{\sqrt{\sigma}}<R<\frac{8}{\sqrt{\sigma}}$.
For each value of $R$ and $L$ we used $10^5$ iterations to thermalize the lattice and then performed $10^6$ measures. Errors were evaluated with the standard blocking  and
jacknife techniques. 

Let us see  in detail the analysis of the $\beta=0.75180$ data.

\subsubsection{Results for $\beta=0.75180$}

The results of the simulations are reported in tab.\ref{tab1}
For this value of $\beta$, the value of $\sigma$ obtained from the scaling function eq.(\ref{powerfit}) is
$\sigma=0.0105255(11)$ (see \cite{Caselle:2007yc}). It is interesting to notice that in this case we also have an estimate for sigma directly obtained in the gauge theory:
$\sigma=0.0105241(15)$ from \cite{Caselle:2004jq}. The fact that
the two values agree within the errors represents a strong crosscheck of the procedure that we adopted
to extract $\sigma$. In the following we shall use the first estimate which is slightly more precise. 
Using this value and the scaling analysis discussed in the previous
section we obtain for $\alpha$ the following prediction $\alpha=2.792~~10^{-5}$. To test this prediction and at the same time the form of the effective string correction
we carefully selected two sets of values of $L$. In the first set we chose three values of $L$. $L=16,20,24$ 
for which the variable $x$ (defined in eq.(\ref{defx})) was tuned so as to make negligible within the errors the sextic
term $x^3/8$, but not negligible the quartic one $x^2/8$ in eq.(\ref{def1}). In the second set we chose 
three values of $L$: $L=10,11,12$ for which instead the sextic term was definitely not
negligible within the errors.

\begin{table}[ht]
\hskip -1cm
\begin{tabular}{| c | c | c | c | c | c | c |} 
\hline
 \  & L = 10 & L = 11 & L = 12 & L = 16 & L = 20 & L = 24  \\
\hline 
R =  8 & N/D & 0.913241(5) & 0.912778(4) & 0.912136(3) & 0.912022(3) &  0.911996(3)\\
\hline
R = 10 & N/D & 0.913300(5) & N/D & N/D & N/D & N/D \\
\hline
R = 12 & 0.914175(7) & 0.913357(5) & 0.912896(4) & 0.912251(3) & N/D & N/D \\
\hline
R = 14 & N/D & 0.913421(5) & N/D & N/D & N/D & N/D \\
\hline
R = 16 & 0.914303(7) & 0.913486(5) & 0.913013(4) & 0.912362(3) & 0.912249(3) & 0.912225(3) \\
\hline
R = 20 & 0.914442(7) & 0.913619(5) & 0.913138(4) & 0.912484(3) &    N/D      &     N/D     \\
\hline
R = 24 & 0.914563(7) & 0.913733(5) & 0.913256(4) & 0.912593(3) & 0.912472(3) & 0.912450(3) \\
\hline
R = 32 & 0.914790(8) & 0.913969(5) & 0.913488(4) & 0.912823(3) & 0.912697(3) & 0.912664(3) \\
\hline
R = 40 & 0.915052(8) & 0.914194(5) & 0.913730(4) & 0.913052(3) & 0.912921(3) & 0.912893(3) \\
\hline
R = 48 & 0.915279(8) & 0.914442(6) & 0.913958(5) & 0.913284(3) & 0.913144(3) & 0.913121(4) \\
\hline
R = 56 & 0.915524(9) & 0.914688(6) & 0.914188(5) & 0.913506(3) & 0.913371(3) & 0.913344(4) \\
\hline
R = 64 & 0.915765(9) & 0.914906(6) & 0.914423(5) & 0.913734(3) & 0.913601(3) & 0.913568(4) \\
\hline 
\end{tabular}
\caption{\small \textit{Values of $\Phi(R,L)$ for $\beta=0.75180$}}
\label{tab1}
\end{table}

For each value of L we fitted the data for $\Phi(R,L)$ according to the law
\eq
\Phi(R,L)=a(L)R+b(L)+c(L)/R
\label{fit10}
\en 
where the term $c(L)/R$ was introduced only for the second set of values of $L$ because it was always compatible with zero in the first set. 
The results of the fits are reported in tab.\ref{tab2}. 

\begin{table}[ht]
\centering %
\begin{tabular}{| c | c | c | c | c | c | c |} 
\hline
\ & $b(L)$ & $a(L)$ & $c(L)$ & $\chi^2$ \\
\hline
L=10 & 0.913840(9)  & 3.016(21)  e-05 & -0.000213(55) & 1.29 \\
\hline
L=11 & 0.913033(10) & 2.944(20)  e-05 & -0.000257(77) & 1.87 \\
\hline
L=12 & 0.912560(7)  & 2.917(13)  e-05 & -0.000145(68) & 0.84 \\
\hline
L=16 & 0.911914(3)  & 2.8472(62) e-05 &   N/D     & 0.81 \\
\hline
L=20 & 0.911795(5)  & 2.8163(86) e-05 &   N/D     & 0.66  \\
\hline
L=24 & 0.911773(3)  & 2.8025(78) e-05  &  N/D          & 1.37  \\
\hline
\end{tabular}
\caption{Results of the fits to eq.(\ref{fit10}) for  $\beta=0.75180$}
\label{tab2}
\end{table}


In the following we shall concentrate on the values of $a(L)$ which as discussed above is the only one which is not affected by boundary corrections.

The values of $a(L)$ extracted from the fits are reported in tab.\ref{tab4} and plotted in fig.1. We analyzed these data in two steps. First 
we fitted the first three values of $a(L)$ (those corresponding to $L=16,20,24$) with the law:
\eq
a(L)=\alpha(1+\gamma x^2)
\en
we found $\alpha=2.7918(17)~10^{-5}$ and  $\gamma= 0.132(7)$ with a very good $\chi^2$. Both these values nicely agree with the predictions $\alpha=2.792~~10^{-5}$
and $\gamma=1/8$. Notice that if we would neglect the scaling correction discussed in the previous section and use eq.(\ref{naive}) to estimate $\alpha$ we would find
$\alpha=2.807~~10^{-5}$ in complete disagreement with the numerical result. 
This gives an idea of the precision of the our estimates and of the need of carefully controlling scaling corrections.

We then fitted the  whole set of data (i.e. including also $L=10,11,12$) with the law:
\eq
a(L)=\alpha(1+\gamma x^2+\delta x^3)
\en
we found again a good reduced $\chi^2$:  $\chi^2_r=0.45$ and the following best fit results:
$$ \alpha=2.796(5)~10^{-5},~~~ \gamma= 0.127(25),~~~ \delta= -0.051(27) $$
The first two values agree again very well with the predictions but the coefficient of the sextic correction, which should be $\delta=1/8$ completely disagrees. 
This can be
well appreciated looking at fig.1
where we plotted the quartic correction (dashed curve), the sextic correction according to the Nambu-Goto effective action (dashed-dotted curve)
and that corresponding to the best fit value of the parameter $\delta$ (dotted curve).

\begin{table}[htpb]
\centering
\begin{tabular}{|c|c|c|c|c|}
  \hline
  $L$ & $a(L)$  &  &  &      \\
  \hline\hline
  10  & 3.017(21)  & 2.792 & 3.137   & 3.481      \\
  11  & 2.943(20)  & 2.792 & 3.027   & 3.222   \\
  12  & 2.917(13)  & 2.792 & 2.959   & 3.074  \\
  16  & 2.847(9)  & 2.792 & 2.845    & 2.865 \\
  20  & 2.816(9)  & 2.792 & 2.814    & 2.819 \\
  24  & 2.802(8)  & 2.792 & 2.802   & 2.804 \\
  \hline
\end{tabular}
\caption{\small \textit{Values of the coefficient $a(L)$ 
for various values of $L$. In the second column we list the results of the simulations performed at $\beta=0.75180$.
In the following columns we 
report the prediction for $a(L)\equiv \alpha A(x)$ at the zeroth, first and second order in the expansion in $x$}}
\label{tab4}
\end{table}

\begin{figure}[htpb]
  \hskip -2cm
  \includegraphics[width=6 in]{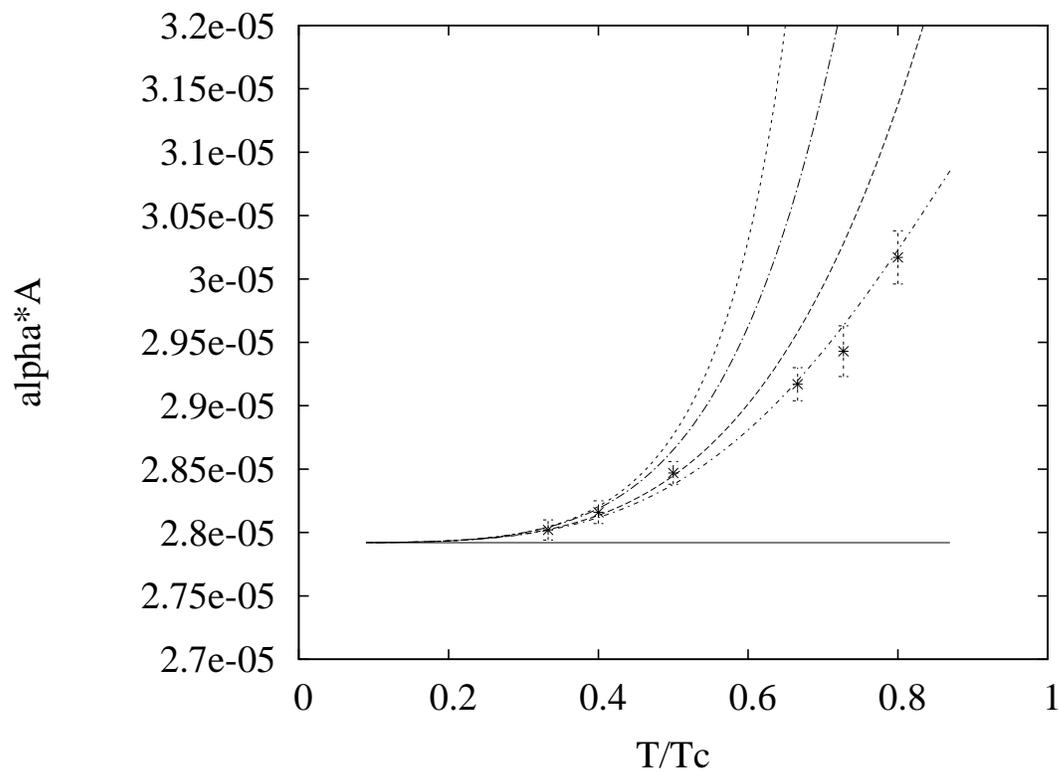}
  \caption{\small \textit{Plot of $\alpha~A(L)$ as a function of $\frac{T}{T_c}=\frac{8}{L}$. The continuous line is the prediction for 
  $\alpha$ (i.e. without effective string corrections. The other curves correspond, from top to bottom to the whole Nambu-Goto prediction, eq(\ref{ng1}), 
  the truncation at the
  sextic order, the truncation at the quartic order and finally the dashed dotted line corresponds to the best fit choice ($\delta= -0.05$) 
  of the sextic coefficient (see the text).
  The points are the results of the simulations in the 3d gauge Ising model
  }}
  \label{fig1}
\end{figure}

\subsubsection{Results for $\beta=0.756427$ and $\beta=0.758266$}
In order to test the scaling behaviour of the effective string corrections we also simulated a value of $L$ corresponding to $T/T_c=0.8$ (i.e. the highest value of $T$ among
those discussed in the previous section) for two other values of $\beta$ nearer to the critical point. More precisely we performed simulations at 
$\beta=0.756427$ on a lattice of size $15\times 192^2$ for values of $R$ in the range $12\leq R\leq 84$ and at 
$\beta=0.758266$ on a lattice of size $20\times 256^2$ for values of $R$ in the range $16\leq R\leq 128$. In order to keep under control the statistical uncertainty we
perfomed in these cases $15\times 10^6$ and  $5\times 10^6$ measures respectively. We report the results of the simulatioins in tab.\ref{tab6} and \ref{tab7}.
Moreover, in order to eliminate spurious effects due to a possible breaking of
the string picture for small values of $R$  we decided to fit only the data in the range $R>\frac{1.5}{\sqrt{\sigma}}$. With this choice the $1/R$ term in the fit was
always compatible with zero. Thus we decided to eliminate it from the fit and perform only two parameters fits (constant plus linear term).

\begin{table}[ht]
\centering %
\begin{tabular}{| c | c |} 
\hline
 \  & L = 15   \\
\hline 
R = 12  &  0.928413(4) \\
\hline
R =  18 &  0.928496(5) \\
\hline
R =  24   & 0.928572(5) \\
\hline
R = 30 & 0.928638(5) \\
\hline
R = 36 &  0.928716(5) \\
\hline
R = 42 &  0.928785(5) \\
\hline
R =48  &  0.928841(5)\\
\hline
R = 54  &  0.928922(5) \\
\hline
R = 60  &  0.928976(5)   \\
\hline
R = 66  &  0.929051(5) \\
\hline
R = 72  &  0.929120(6)  \\
\hline
R = 78  &  0.929187(6)  \\
\hline
R =84  &  0.929257(6)  \\
\hline
R = 90&   0.929335(7) \\
\hline
R = 96  & 0.929387(6)   \\
\hline 
\end{tabular}
\caption{\small \textit{Values of $\Phi(R,L)$ for $\beta=0.756427$}}
\label{tab6}
\end{table}

\begin{table}[ht]
\centering %
\begin{tabular}{| c | c |} 
\hline
 \  & L = 20   \\
\hline 
R = 16 & 0.934812(8) \\
\hline
R = 24 & 0.934874(8)  \\
\hline
R = 32 & 0.934907(8) \\
\hline
R =  40 & 0.934962(7) \\
\hline
R =  48 & 0.935019(8) \\
\hline
R = 56 & 0.935045(8) \\
\hline
R = 64 & 0.935090(6) \\
\hline
R = 72 & 0.935149(8) \\
\hline
R = 80 & 0.935189(8) \\
\hline
R = 88 & 0.935240(9) \\
\hline
R = 96 & 0.935279(8) \\
\hline
R = 104 & 0.935344(6) \\
\hline
R = 112 & 0.935398(11) \\
\hline
R = 120 & 0.935429(12) \\
\hline
R = 128 & 0.935477(10) \\
\hline 
\end{tabular}
\caption{\small \textit{Values of $\Phi(R,L)$ for $\beta=0.758266$}}
\label{tab7}
\end{table}

\begin{figure}[htpb]
  \hskip -2cm
  \includegraphics[width=6 in]{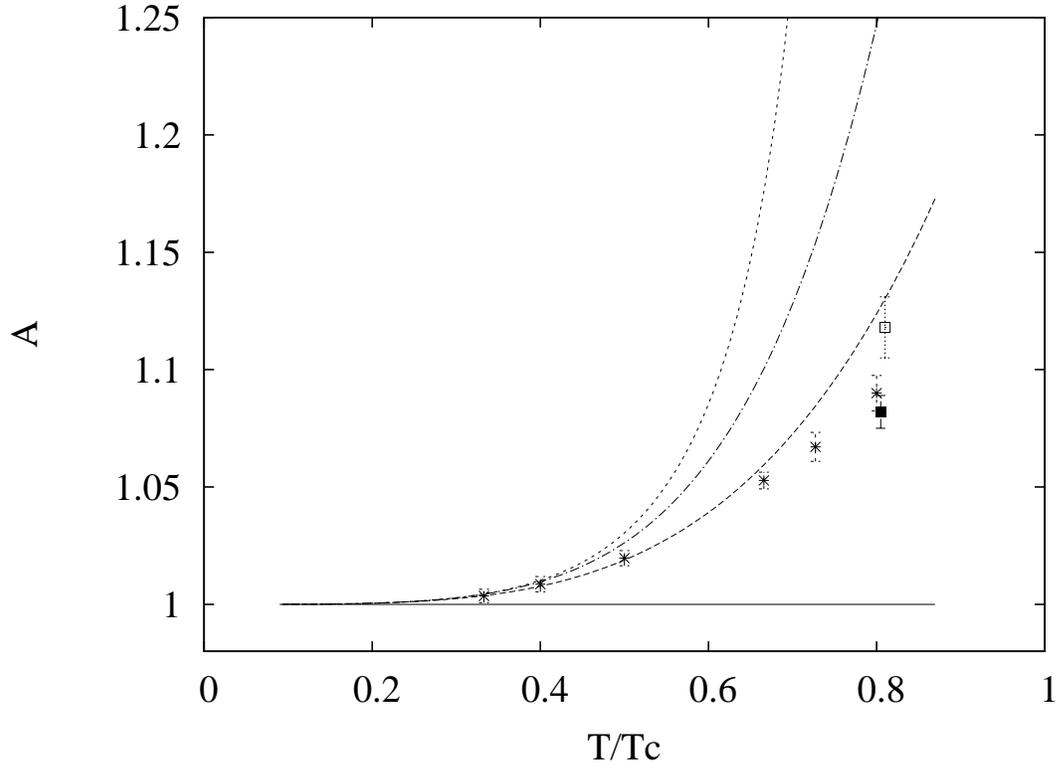}
  \caption{\small \textit{Plot of $A(L)$ as a function of $\frac{T}{T_c}$. As in fig.1 the continuous line is the prediction for 
  $\alpha$ (i.e. without effective string corrections. The other curves correspond, from top to bottom to the whole Nambu-Goto prediction, eq(\ref{ng1}), 
  the truncation at the
  sextic order and the truncation at the quartic order. 
  The points are the results of the simulations (see text). Stars denote the data at $\beta=0.75180$, the black square is the value for $\beta=0.756427$, $T=0.8 T_c$ 
  and and the open square the point at $\beta=0.758266$
  }}
  \label{fig2}
\end{figure}

The results are reported in tab.\ref{tab5} and fig.\ref{fig2} 
where we have plotted the function $A(x)$ defined in eq.(\ref{def1})as a function of the finite temperature $T/T_c$. Since both quantities
are adimensional we may plot different values of $\beta$ on the same graph. In fig.\ref{fig2} we also reported the Nambu-Goto prediction for $A(x)$ (dotted line) together
with the value truncated at the fourth order (dashed line) and at the sixth order (dashed dotted line). 

It is easy to see that all the points lie $below$ the fourth order line (i.e. they are compatible with a negative value of $\delta$) and  
show a slight tendency to move toward the value $\delta=0$ as $\beta$ moves toward the continuum limit. In all the cases the data are  incompatible with
$\delta=1/8$. This test seems to suggest that the
deviations that we observe from the universal sixth order term 
is not due to the effect of an irrelevant operator but, at least in the range of values that we studied, seems to
hold up to the continuum limit.

\begin{table}[htpb]
\centering
\begin{tabular}{|c|c|c|c|c|c|}
  \hline
  $\beta$ & $T/T_c$ & $A(x)$  &  &  &      \\
  \hline\hline
   $0.751800$ &0.333  &1.0036(29)  & 1.0037 & 1.0044   & 1.0045 \\
   $0.751800$ &0.4  & 1.0086(32)  & 1.0077& 1.0097    & 1.0102 \\
   $0.751800$ &0.5  & 1.020(3)  & 1.0189 & 1.0262    & 1.0304 \\
   $0.751800$ &0.667  & 1.052(4)  & 1.0597 & 1.1009   & 1.1773  \\
   $0.751800$ &0.727  & 1.067(6)  & 1.0845 & 1.1540   & 1.3967   \\
   $0.751800$ & 0.8  & 1.090(8)  & 1.1237 & 1.2468   & 7.0472      \\
   $0.756427$ &0.8 & 1.082(7)  & 1.1237 & 1.2468  & 7.0472 \\
   $0.758266$ &0.8  & 1.118(13)  & 1.1237 &  1.2468  & 7.0472 \\
  \hline
\end{tabular}
\caption{\small \textit{Values of the coefficient $A(x)$ 
for various values of $\beta$ and $T/T_c$. In the third column we list the results of the simulations.
In the following two columns we 
report the prediction for $A(x)$ truncated at the first and second order
 in the expansion in $x$}. In the last column the all orders result for $A(x)$.}
\label{tab5}
\end{table}

\section{Conclusions}
In this paper we discussed a new strategy to test effective string predictions. Instead of looking at the interquark potential we studied 
the flux distribution in presence of
the quark-antiquark pair. The major consequence of this choice is that the leading effective string correction (the L\"uscher term) disappears thus allowing to disentangle
higher order corrections. We apply the method to the 3d gauge Ising model and find that while the first subleading correction (the one related to quartic terms in the
string Lagrangian) 
agrees very well with the one predicted by the Nambo-Goto model, the subsequent term 
(due to sixth order terms in the Lagrangian) is definitely different and seems to have the opposite
sign or at most to be compatible with zero. This observation agrees with a set of similar results obtained in these last years in the 3d gauge Ising model 
looking at various different physical observables
ranging from the interquark potential to the effective string width to the interface free energy of the dual spin Ising
model\cite{chp03,Caselle:2004jq,Caselle:2007yc,Allais:2008bk}. These tests were all
 supporting, even if only at a qualitative level, a value for the sextic correction different
from the Nambu-Goto one and fully compatible with the one that we find here.

Our result 
 is rather puzzling in view of the recent proof of the universality of the Nambu-Goto action up to this order in $d=3$~\cite{Aharony:2009gg,ak,af}. We see a few 
possible explanations for this disagreement.
\begin{itemize}
\item It could be due to an irrelevant operator and thus it would disappear in the continuum limit. While we cannot exlude in principle this possibility the scaling test
discussed at the end of the previous section suggests that this is not the case. 

\item The effect that we observe could be due to a compensation between the sixth order and all the remaining higher order terms which are not constrained by universality.
While this is certainly possible, it would require a rather strong fine tuning of  the coefficients of all the higher order terms. If this is the explanation, 
then it would be important to understand the reason of such a cancellation, which most probably doesn't occur by chance but would be the signature of 
some underlying constraint or symmetry.
  
\item It could be due to the fact that 3d gauge Ising model does not admit a weakly coupled effective string description. This  limit 
is a basic assumption  of \cite{lw04,Aharony:2009gg,Billo:2005iv} and amounts to ask that the partition function of the string describing a particular surface (say
the cylindric surface connecting the two Polyakov loops) can be written as a sum of single string states propagating along the surface. 
The fact that this argument could be relevant for the 3d gauge Ising model is also supported by the intuitive observation that a gauge theory based on the 
$Z_2$ group (and even more the 
percolation gauge theory discussed in \cite{Giudice:2009di} ) is indeed the farthest 
possible choice with respect to the large $N$ limit of the SU(N) gauge theory which is known to behave as a weakly coupled string theory. If this is the reason it
would be interesting to understand why the effective string description works instead so well up to the quartic order.

\end{itemize}

\vskip1.0cm {\bf Acknowledgements.}
 We  warmly thank M.Bill\'o and  F. Gliozzi for useful discussions and suggestions. 
 M.C. would like to thank all the partecipants of  
 the {\it Confining Flux Tubes and Strings} Workshop at the ECT*, 
Trento for several useful discussions 
which partially stimulated the present work. This research was supported in
part by the European Community - Research Infrastructure Action under the
FP7 "Capacities'' Specific Programme, project "HadronPhysics2''.

\end{document}